\documentclass[11pt]{article}
\usepackage[margin=2cm]{geometry}

\usepackage{graphicx} 

\usepackage{amssymb}
\usepackage{amsmath}
\usepackage{braket}
\usepackage{indentfirst}
\usepackage{hyperref}
\usepackage{authblk} 
\usepackage{multirow} 
\usepackage[T1]{fontenc} 

\begin{document}

\title{Waterlike density anomaly in fermions}

\author[1]{Marco A. Habitzreuter}
\author[2]{Eduardo O. Rizzatti}
\author[1]{Marcia C. Barbosa\thanks{\href{mailto:marcia.barbosa@ufrgs.br}{marcia.barbosa@ufrgs.br}}}
\affil[1]{\footnotesize Instituto de Física - Universidade Federal do Rio Grande do Sul, Porto Alegre, 91501-970, Rio Grande do Sul, Brazil.}
\affil[2]{\footnotesize Laboratory of Atomic and Solid State Physics - Cornell University, Ithaca, 14853, New York, USA.}

\date{September 21, 2023}

\maketitle

\begin{abstract}
In this work we explore the one-dimensional extended Hubbard model as a fluid system modelling liquid phases of different densities. This model naturally displays two length scales of interaction, which are connected with waterlike anomalies. We analyze the density anomaly as a function of the model parameters, namely the hopping, on-site and first neighbor interactions. We show that this anomaly is present for a wide range of model parameters and is connected to a ground-state liquid-liquid critical point.
\end{abstract}

\section{Introduction}
\label{sec:introduction}

At first glance, the water molecule might seem uninteresting due to its simple molecular formula. But on a closer look, we see a myriad of surprises. The importance of water as a resource to life makes the study of its properties fundamental to many areas of science. For instance, its large heat capacity is related to the control of temperature in our environment~\cite{franks2000water} and the properties of supercooled water influence how radiation from the Sun is reflected or absorbed~\cite{debenedetti1996metastable}.

Arguably the most known anomaly is the temperature of maximum density (TMD). Unlike most liquids, water exhibits a peculiar behavior where its volume decreases as the temperature increases within a specific temperature range. Specifically, at a pressure of 1 atm, the density of liquid water increases from 0\textdegree C up to 4\textdegree C, where it reaches a maximum~\cite{kell_density_1975}. This property holds significant importance for the preservation of aquatic life in certain cold lakes, as it allows denser water to remain in its liquid state at the lake's bottom, while the surface freezes.

This peculiar density behavior is elucidated by the interplay between two distinct structures: an open arrangement and a compact arrangement of water molecules, resulting in a low-density and high-density liquid, respectively. Experimental studies in the 1980's indicated a transition between high-density and low-density amorphous ice~\cite{mishima_apparently_1985}. Subsequently, computational models were employed to extend this transition from high-density to low-density liquid~\cite{scp_stanley}. Due to hydrogen bonding, a network connects water molecules at specific angles and distances, fostering an open structure at low temperatures, which corresponds to the low-density liquid state. As the temperature increases, these bonds in the liquid are disrupted, causing the structure to become more compact and resulting in a higher density network. The coexistence line between these two liquid phases ends in a liquid-liquid critical point. This microscopic mechanism gives rise to entropy and volume fluctuations, which are connected to response functions. As temperature rises, these fluctuations diminish. For example, the thermal expansion coefficient, which relates to volume and entropy fluctuations, exhibits negative values at low temperatures due to the anti-correlation between these fluctuations, and it diverges at the critical point. As temperature increases, the fluctuations decrease, resulting in a decrease in the thermal expansion coefficient until it reaches a point where $\alpha_P = 0$, leading to a density anomaly~\cite{debenedetti_supercooled_2003,franzese2007widom}.

Experimental verification of the hypothesis that water anomalies arise from the competition of two length scales, which also gives rise to a liquid-liquid critical point, poses a considerable challenge. The difficulty lies in the fact that the proposed liquid-liquid critical point appears to be situated within a pressure and temperature range where ice nucleation occurs rapidly. Consequently, evidence for the existence of two liquid phases appears only in out-of-equilibrium experiments~\cite{kim_experimental_2020}.

The relevance of and interplay between two length scales and different liquid structures led to the introduction of purely theoretical models with two length scales modeling the interactions. Simulations and analytic calculations of these system indicated the presence of density, diffusion and structural anomalies similar to the anomalies observed in liquid water~\cite{alan2008, scala2001}, making these so-called core-softened potentials an object of extensive study~\cite{prestipino2012density,buldyrev2002models,fomin2013silicalike,franzese2001generic}. The concept was tested further, with analysis of systems interacting by potentials with multiple length scales showing multiple regions of density anomaly~\cite{rizzatti2018core,barbosa2013multiple}.

The proposed liquid-liquid transition is not exclusive to water. Simulations of carbon, silicon and silica~\cite{glosli1999liquid,sastry2003liquid,saika2005simulated} present density anomaly and experiments with fluid phosphorous~\cite{monaco2003nature} show evidences of double criticality. In fact, density anomalies were even measured in liquid helium-3~\cite{boghosian1966density} and helium-4~\cite{niemela1995density}.

In this perspective, Rizzatti and collaborators~\cite{rizzatti_waterlike_2019,rizzatti2020quantum} recently proposed that the link between two length scales, anomalies and double criticality could be observed in a bosonic system interacting in optically trapped ultracold gases. The authors studied a bosonic lattice system represented by the Bose-Hubbard model with an in-site repulsive interaction. They showed that the competition between the chemical potential and the in-site repulsion creates two main regions of density anomaly: one in the vicinity of the regular fluid regime and the other in the superfluid phase. The results also showed multiple regions of density anomaly closing on the critical phase transition.

On the other hand, condensed matter physicists have used the Fermi-Hubbard model for decades to research phenomena in strongly correlated electron systems in a variety of lattice systems. Many extensions to this model were also proposed, such as multiple bands, nearest and next-nearest neighbor interaction, pair hopping and next nearest neighbor hopping, among others~\cite{Dutta_2015,PhysRevB.44.770,beni1974thermodynamics,PhysRevB.35.3359,robaszkiewicz_superconductivity_1999,PhysRevB.46.5496}. In the case of fermions this model is usually used with a repulsive on-site interaction in connection with the Coulomb repulsion. However, an attractive interaction might also be produced through Feshbach resonances~\cite{RevModPhys.82.1225}. In the last two decades this attractive interaction gained attention due to the experimental realization of the BEC-BCS Crossover~\cite{greiner2003emergence}: a high attraction between fermions can lead to a bosonic-like behavior.

Since waterlike anomalies are present in the Bose-Hubbard model, it is natural to question if they would also be present with fermions. If so, the hypothesis that competing interactions, anomalies and criticality are connected would gain more evidences of its universality. In particular, the extended Hubbard model displays two length scales of interaction, similar to the potentials generating anomalies in molecular dynamics simulations for water. Hence, the study of thermodynamic anomalies in cold-atoms systems is both interesting on its own, but also could provide hints to water behavior. 

In this work, we study the presence a waterlike density anomaly in a fermionic system. Particularly, we explore the extended Hubbard model in one dimension. This model describes spin $\frac{1}{2}$ fermions in a lattice with a hopping probability, local (on-site), as well as nearest neighbors interactions. Considering spin $\frac{1}{2}$ fermions, four possible occupations are allowed for each site: $\ket{0}$ (empty), $\ket{\uparrow}$ (occupied by spin up), $\ket{\downarrow}$ (occupied by spin down) and $\ket{\uparrow\downarrow}$ (doubly occupied with opposite spins). This model has been numerically studied through mean-field~\cite{micnas_superconductivity_1990}, DMRG~\cite{iemini_entanglement_2015}, self-energy functionals~\cite{aichhorn_charge_2004}, Quantum Monte Carlo~\cite{yao_determinant_2022} and other techniques. An exact solution to this model is known only in the atomic limit~\cite{mancini_extended_2005}. We treat the thermodynamics of these fermions using a two site exact diagonalization scheme. Our analysis demonstrates that such system presents density anomalies for a wide range of parameters.

This paper is structured as follows. In Section~\ref{sec:model} we introduce the model and the approximation method to treat its thermodynamics. Section~\ref{sec:results} shows the results in two scenarios: attractive and repulsive on-site interactions, for generic values of nearest neighbor interactions, both in the atomic limit and with a finite hopping. We end with our conclusions and perspectives for future work in Section~\ref{sec:conclusion}.

\section{Model and Methods}
\label{sec:model}

We consider the single-band extended Hubbard model given by the Hamiltonian
\begin{equation}
    \mathcal H = -t \sum_{\braket{i, j}, \sigma} \left(c^{\dagger}_{i, \sigma} c_{j, \sigma} + \text{h.c} \right) + U \sum_{i} n_{i, \downarrow} n_{i, \uparrow} + V\sum_{\braket{i, j}} n_i n_j - \mu \sum_{i} n_{i} \;,
\end{equation}
where
\begin{align}
    n_{i, \sigma} &= c^{\dagger}_{i, \sigma} c_{i, \sigma} \;,\\
    n_i &= n_{i, \downarrow} + n_{i, \uparrow} \;,
\end{align}
and $c^{\dagger}_{i, \sigma}$ ($c_{i, \sigma}$) is the creation (annihilation) operator for an electron of spin $\sigma$ on a site $i$ of the lattice. The parameter $t$ denotes the hopping between different Wannier sites and $U$ ($V$) is the intra-site (inter-site) interaction. The notation $\braket{i, j}$ indicates that we are summing only over the first neighbor sites.

In the following analysis we consider a one-dimensional chain in a pairwise approximation: for $N$ sites, we have $N/2$ pairs, and we assume that each pair is independent. That is, $Y = Y^{N/2}_{\text{pair}}$, where $Y$ is the Grand Canonical partition function and $Y_{\text{pair}}$ is the Grand Canonical partition function for a single dimer. The Grand Potential for the system is
\begin{equation}
    \Omega = -k_B T \ln Y
    \label{eq:grandPotential}
\end{equation}
and the density is given by $\rho = -\frac{1}{\mathcal V} \left(\frac{\partial \Omega}{\partial \mu}\right)_{T, \mathcal V}$, where $\mathcal V$ is the volume.

\subsection{The atomic limit and the introduction of a new notation}

In the atomic limit ($t = 0$), consider a pair of adjacent sites $\braket{i, j}$. The corresponding pair Hamiltonian takes on the following form:
\begin{equation}
    H^{0}_{pair} = U(n_{i,\downarrow}n_{i,\uparrow}+n_{j,\downarrow}n_{j,\uparrow}) +2 V n_in_j - \mu(n_i+n_j)\;,
    \end{equation}
which is diagonal on the occupation number basis. The eigenvalues can be expressed as
\begin{equation}
    E^{0}_{(m,n)} = U(\delta_{m,2}+\delta_{n,2}) +2 V m n - \mu(m+n) \;,
\end{equation}
where $m$ and $n$ denotes the occupation of the sites in the pair, with $0 \leq n \leq m\leq 2$. In this notation the pair $(0,0)$ corresponds to $\ket{0,0}$, which is the empty state. The pair $(1,0)$ represents 4 states with one singly occupied site and the other empty: $\ket{\downarrow,0}$, $\ket{0,\downarrow}$, $\ket{\uparrow,0}$, $\ket{0,\uparrow}$, and so on. The eigenvalues and the respective eigenstates are summarized in Table~\ref{tab:eigenstates_atomic_limit}. Therefore, the partition function of the pair is given by
\begin{equation}
    Y_{pair} = \sum_{m \geq n} g_{(m,n)} e^{-\beta E^{0}_{(m,n)}} \;,
    \label{eq:2sc_partition_function}
\end{equation}
where $g_{(m,n)}$ is the degeneracy associated to each pair $(m,n)$.

\begin{table}[h]
    \center
    \begin{tabular}{c|c|c|c|c}
        N & Pair $(m,n)$ & Eigenstate & Eigenvalue & Degeneracy  \\
        \hline
        \hline
        $0$ & $(0,0)$ & $\ket{0,0}$ & $E^0_{(0,0)}=0$ & $g_{(0,0)}=1$ \\
        \hline
        \multirow{4}{*}{$1$} &  \multirow{4}{*}{$(1,0)$} & $\ket{\downarrow,0}$ & \multirow{4}{*}{$E^0_{(1,0)}=-\mu$} & \multirow{4}{*}{$g_{(1,0)}=4$} \\
        & & $\ket{0,\downarrow}$ & &\\
        & & $\ket{\uparrow,0}$ & &\\
        & & $\ket{0,\uparrow}$ & &\\
        \hline
        \multirow{6}{*}{$2$} &  \multirow{2}{*}{$(2,0)$} & $\ket{\uparrow\downarrow,0}$ & \multirow{2}{*}{$E^0_{(2,0)}=U-2\mu$} & \multirow{2}{*}{$g_{(2,0)}=2$} \\
        & & $\ket{0,\uparrow\downarrow}$ & &\\
        \cline{2-5}
        &  \multirow{4}{*}{$(1,1)$} & $\ket{\downarrow,\downarrow}$ & \multirow{4}{*}{$E^0_{(1,1)}=2V-2\mu$} & \multirow{4}{*}{$g_{(1,1)}=4$} \\
        & & $\ket{\uparrow,\uparrow}$ & &\\
        & & $\ket{\downarrow,\uparrow}$ & &\\
        & & $\ket{\uparrow,\downarrow}$ & &\\
        \hline
        \multirow{4}{*}{$3$} &  \multirow{4}{*}{$(2,1)$} & $\ket{\uparrow\downarrow,\downarrow}$ & \multirow{4}{*}{$E^0_{(2,1)}=U+4V-3\mu$} & \multirow{4}{*}{$g_{(2,1)}=4$} \\
        & & $\ket{\downarrow,\uparrow\downarrow}$ & &\\
        & & $\ket{\uparrow \downarrow,\uparrow}$ & &\\
        & & $\ket{\uparrow,\uparrow \downarrow}$ & &\\
        \hline
        $4$ & $(2,2)$ & $\ket{\uparrow \downarrow,\uparrow \downarrow}$ & $E^0_{(2,2)}=2U+8V-4\mu$ & $g_{(2,2)}=1$ \\
        \hline
    \end{tabular}
    \caption{Description of the eigenstates and eingenvalues in the atomic limit. The pair $(m,n)$ labels a phase of energy $E^0_{(m,n)}$ and degeneracy $g(m,n)$, containing $m$ fermions on one site and $n$ on the other, where $N=m+n$.}
    \label{tab:eigenstates_atomic_limit}
\end{table}

\subsection{Finite hopping}
Considering a finite hopping ($t>0$),  the pair Hamiltonain is given by
\begin{equation}
    H_{pair} =  U(n_{i,\downarrow}n_{i,\uparrow}+n_{j,\downarrow}n_{j,\uparrow}) +2 V n_in_j - \mu(n_i+n_j)
    -t(c^{\dagger}_{i, \downarrow}c_{j, \downarrow} + c^{\dagger}_{j, \downarrow}c_{i, \downarrow} + c^{\dagger}_{i, \uparrow}c_{j, \uparrow} + c^{\dagger}_{j, \uparrow}c_{i, \uparrow})\;.
    \label{eq:Hpair}
\end{equation}
After its diagonalization, the resulting set of eigenstates and eigenvalues of Eq.~\ref{eq:Hpair} is shown in Table~\ref{tab:eigenstates_finite_hopping}. The notation in the table notation was inspired by the atomic limit results. The upper bars denote states with higher energy, compared to the respective configurations without bars. We clearly note that the inclusion of the hopping lifts many degeneracies seen in the atomic limit (this can be directly read from the table, were we have more eigenvalues and smaller values for $g$).

\renewcommand{\arraystretch}{1.3}
\begin{table}[h]
    \center
    \begin{tabular}{ c | c | c | c | c }
        N & Pair $(m,n)$ & Eigenstate & Eigenvalue & Degeneracy \\
        \hline
        \hline
        $0$ & $(0,0)$ & $\ket{0,0}$ & $E^0_{(0,0)}=0$ & $g_{(0,0)}=1$ \\
        \hline
        \multirow{4}{*}{$1$} & \multirow{2}{*}{$(1,0)$} & $\ket{\downarrow,0}+\ket{0,\downarrow}$ & \multirow{2}{*}{$E_{(1,0)}=E^0_{(1,0)} - t$} & \multirow{2}{*}{$g_{(1,0)}=2$} \\
        & & $\ket{\uparrow,0}+\ket{0,\uparrow}$ & & \\
         \cline{2-5}
        & \multirow{2}{*}{$\overline{(1,0)}$} & $\ket{\downarrow,0}-\ket{0,\downarrow}$ & \multirow{2}{*}{$E_{\overline{(1,0)}}=E^0_{(1,0)} + t$} & \multirow{2}{*}{$g_{\overline{(1,0)}}=2$} \\
        & & $\ket{\uparrow,0}-\ket{0,\uparrow}$ & & \\
        \hline
        \multirow{6}{*}{$2$} & \multirow{3}{*}{$(1,1)$} & $\ket{\uparrow,\uparrow}$ & \multirow{3}{*}{$E_{(1,1)}=E^0_{(1,1)}$} & \multirow{3}{*}{$g_{(1,1)}=3$} \\
        & & $\ket{\downarrow,\downarrow}$ & &\\
        & & $\ket{\downarrow,\uparrow}-\ket{\uparrow,\downarrow}$ & &\\
        \cline{2-5}
        & $(2,0)$ & $\ket{\uparrow \downarrow,0}-\ket{0,\uparrow \downarrow}$ & $E_{(2,0)}=E^0_{(2,0)}$ & $g_{(2,0)}=1$ \\
        \cline{2-5}
        & $(2,0)\bigoplus(1,1)$ & $\alpha\ket{\uparrow,\downarrow}+\beta\ket{\downarrow,\uparrow}+\gamma\ket{\uparrow \downarrow,0}+\delta\ket{0,\uparrow \downarrow}$ & $E_{(2,0)\bigoplus(1,1)}=\lambda$ & $g_{(2,0)\bigoplus(1,1)}=1$ \\
        \cline{2-5}
        & $\overline{(2,0)}\bigoplus \overline{(1,1)}$ & $\bar{\alpha}\ket{\uparrow,\downarrow}+\bar{\beta}\ket{\downarrow,\uparrow}+\bar{\gamma}\ket{\uparrow \downarrow,0}+\bar{\delta}\ket{0,\uparrow \downarrow}$ & $E_{\overline{(2,0)}\bigoplus\overline{(1,1)}}=\bar{\lambda}$ & $g_{\overline{(2,0)}\bigoplus\overline{(1,1)}}=1$ \\
        \hline
        \multirow{4}{*}{$3$} &  \multirow{2}{*}{$(2,1)$} & $\ket{\uparrow\downarrow,\downarrow}+\ket{\downarrow,\uparrow\downarrow}$ & \multirow{2}{*}{$E_{(2,1)}=E^0_{(2,1)}-t$} & \multirow{2}{*}{$g_{(2,1)}=2$} \\
        & & $\ket{\uparrow\downarrow,\uparrow}+\ket{\uparrow,\uparrow\downarrow}$ & &\\
        \cline{2-5}
        &  \multirow{2}{*}{$\overline{(2,1)}$} & $\ket{\uparrow\downarrow,\downarrow}-\ket{\downarrow,\uparrow\downarrow}$ & \multirow{2}{*}{$E_{\overline{(2,1)}}=E^0_{(2,1)}+t$} & \multirow{2}{*}{$g_{\overline{(2,1)}}=2$} \\
        & & $\ket{\uparrow\downarrow,\uparrow}-\ket{\uparrow,\uparrow\downarrow}$ & &\\
        \hline
        $4$ & $(2,2)$ & $\ket{\uparrow \downarrow,\uparrow \downarrow}$ & $E_{(2,2)}=E^0_{(2,2)}$ & $g_{(2,2)}=1$ \\
        \hline
    \end{tabular}
    \caption{Description of the eigenstates and eingenvalues for finite hopping. In the table, $2\lambda = E^0_{(1,1)}+E^0_{(2,0)} - \sqrt{(E^0_{(1,1)} - E^0_{(2,0)})^2+16t^2})$ and $2\bar{\lambda} = E^0_{(1,1)}+E^0_{(2,0)} + \sqrt{(E^0_{(1,1)} - E^0_{(2,0)})^2+16t^2}$.}
    \label{tab:eigenstates_finite_hopping}
\end{table}

\subsection{Numerical calculations}
Given the list of eigenenergies and the corresponding degeneracy, any thermodynamic quantity can be calculated from the Grand Potential given by Equation~\ref{eq:grandPotential}. We rescale the quantities with an energy unit $\varepsilon_r$ and a length unit $v_0$. Hence,
\begin{align}
    t^{*} &\equiv t/\varepsilon_r, \\
    U^{*} &\equiv U/\varepsilon_r, \\
    V^{*} &\equiv V/\varepsilon_r, \\
    \mu^{*} &\equiv \mu/\varepsilon_r, \\
    \rho^{*} &\equiv \rho v_0, \\
    T^{*} &\equiv k_B T/\varepsilon_r,\\
    P^{*} &\equiv P v_0 /\varepsilon_r,
\end{align}
where $\rho$ is the density per site, $P$ is the pressure and $k_B$ is the Boltzmann constant. For simplicity, we omit the $^{*}$ superscript from now on. In physical terms, $\varepsilon_r$ can be understood as the recoil energy, used as an energy scale in optical lattices experiments, while $v_0$ is the volume of the unit cell of the lattice.

\section{Results}
\label{sec:results}

\begin{figure}
    \centering
    \includegraphics[width=0.4\textwidth]{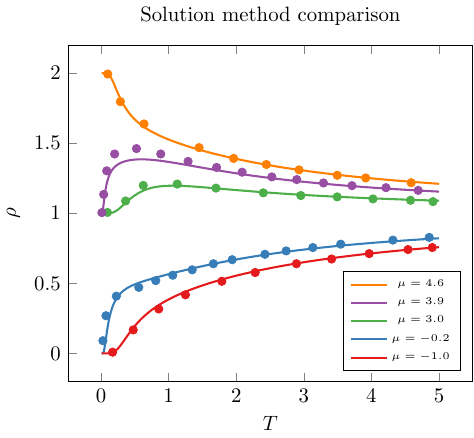}
    \caption{Density as a function of temperature for various values of chemical potential with two solution methods. Solid curves are calculated with the pair approximation, while the points are extracted from Mancini~\cite[Fig. 5]{mancini_extended_2005}.}
    \label{fig:rho_x_T_const_mu_comparison}
\end{figure}

In Figure~\ref{fig:rho_x_T_const_mu_comparison} we plot in solid curves the density as a function of temperature for fixed values of the chemical potential as solid lines. In order to validate the approximation, we compare our results with the values extracted  from the exact atomic limit solution by Mancini et al.~\cite{mancini_extended_2005}, using the formalism of equations of motion. There is quantitative agreement between most of the solid lines and the circles. The larger discrepancy happens for $\mu = 3.9$ and $T \approx 0.7$: the density calculated with the two-site approximation is about 5\% smaller.

In our two-site system, the ground-state phase diagram can be determined by the lowest energy eigenvalue $E^{0}_{(m',n')}$ given a set of parameters $\mu, U, V$:
\begin{equation}
    E^{0}_{(m',n')} = \min_{m,n} \{E^{0}_{(m,n)}\} \;.
\end{equation}
The phase boundaries are given by the crossing between different energy hyperplanes. For instance, for the coexistence between $(0,0)$ and $(1,0)$, we have
\begin{equation}
    E^0_{(0,0)} = E^0_{(1,0)} \;\;\; \iff \;\;\; 0=-\mu  \;\;\; \iff \;\;\; \boxed{\mu =0}
\end{equation}
For the coexistence between $(1,0)$ and $(1,1)$, we get
\begin{equation}
    E^0_{(1,0)} = E^0_{(1,1)} \;\;\; \iff \;\;\; -\mu = 2V- 2\mu \;\;\; \iff \;\;\;  \boxed{V = \mu}
\end{equation}
and so on. Collecting all solutions, we draw the phase diagram for two qualitatively distinct cases in the atomic limit: $U \leq 0$ and $U > 0$. In the next subsections we discuss these diagrams and the density anomaly regions, which appear in both cases. The effects of changing $t$, $U$ and $V$ on the density anomaly are also addressed.

\subsection*{Case 1: $U \leq 0$}

\begin{figure}
    \centering
    \includegraphics[width=0.8\textwidth]{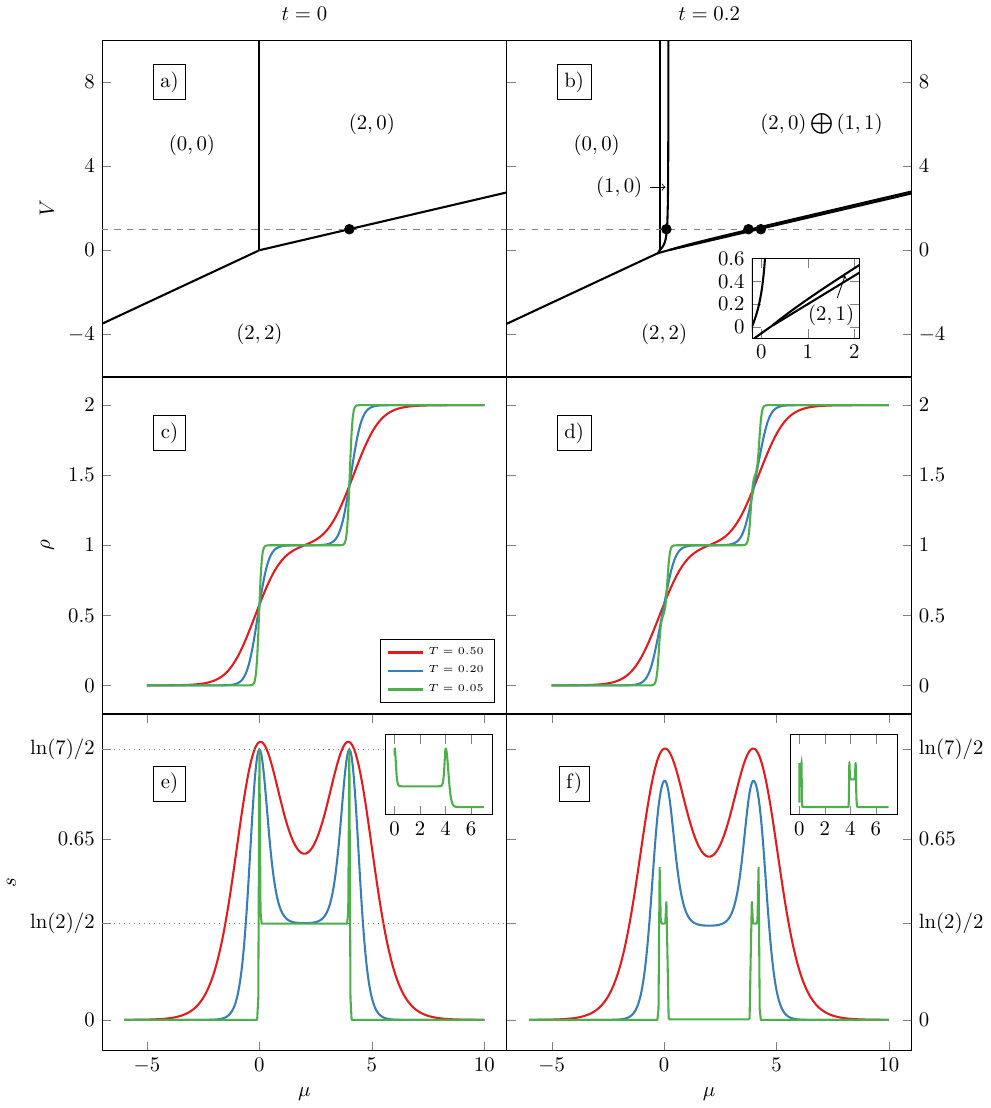}
    \caption{Panels a) and b) portray the ground state phase diagram at $U=0$. The black dots are explained in Figure~\ref{fig:2sc_rho_x_T_const_P_U0_V1}. Panels c) to f) show the density and entropy at a fixed value of $V = 1$ (dashed line) for various temperatures. The left column corresponds to the atomic limit $t=0$ while the right column refers to $t=0.2$. The inset in panel b) zooms in the $0 < \mu < 2$ range to highlight the small $(2, 1)$ phase. Insets in the last two panels are the entropy at $T = 0.01$ as a function of pressure. The same color scheme is used in all plots.}
    \label{fig:2sc_U0_V1}
\end{figure}

The ground state phase diagram for this case is shown in panels a) and b) of Figure~\ref{fig:2sc_U0_V1}, allowing us to identify three phases in the atomic limit: $(0, 0)$, $(2, 0)$ and $(2, 2)$, as identified in Table~\ref{tab:eigenstates_atomic_limit}. In this extended $(\mu, V, T = 0, U = 0)$ phase diagram, continuous lines are first-order transitions between these phases, and they can coincide at a triple point, located at $(\mu = U/2, V = 0)$ in the atomic limit. With a finite hoping amplitude, two other phases appear in the ground state: $(1,0)$ and $(2,1)$. Their sizes, which are considerably smaller than the others, decrease as $U$ becomes negatively large.

The density profile for a fixed value of $V$ is shown in panels c) and d) of Figure~\ref{fig:2sc_U0_V1} for different temperatures. In the limit $T \rightarrow 0$, the density becomes a discontinuous function of the chemical potential, signaling first order phase transitions between states of different densities. These ground state phase transitions occur at critical values of chemical potential $\mu$ such that the energy of each configuration is the same. In this system only ground state phase transitions are allowed, since we are working in one dimension and the interactions have a finite range. As we will show, this transition affects finite temperature behavior.

The entropy per site is determined by the Grand Potential according to
\begin{equation}
    s= -\frac{1}{N}\frac{\partial \Omega}{\partial T} = \frac{1}{2}k_B\left(-\frac{\beta}{Y_{pair}} \frac{\partial Y_{pair}}{\partial \beta}+ \ln Y_{pair} \right) \;.
\end{equation}
At low temperatures $\beta \rightarrow \infty$ and away from the critical points $\mu \neq \mu_c$, the partition function is dominated by the following term
\begin{equation}
    Y_{pair} \approx g_{(m',n')} e^{-\beta E^{0}{(m',n')}} 
\end{equation}
where $E^{0}_{(m',n')} =\min_{m,n} \{E^{0}_{(m,n)}\}$ given $\mu, U, V$. Then,
\begin{equation}
    \frac{\partial Y_{pair}}{\partial \beta} \approx - g_{(m',n')}E^{0}_{(m,n)} e^{-\beta E^{0}_{(m',n')}}
\end{equation}
and the entropy assumes the simplified form
\begin{equation}
    s \approx \frac{1}{2}k_B\left(\beta  E^{0}_{(m',n')} + \ln \left[ g_{(m',n')} e^{-\beta E^{0}_{(m',n')}} \right]\right) = \frac{1}{2}k_B \ln  g_{(m',n')} \;.
\end{equation}
For example, the phase $(2,0)$ yields a residual entropy of $\frac{1}{2}k_B \ln  2$ while $(1,1)$ produces $\frac{1}{2}k_B \ln  4$. This result explains the plateaus observed in the $s$ versus $\mu$ plots near the ground state as shown in panels e) and f) of Figure~\ref{fig:2sc_U0_V1}. They appear due to the degeneracy of a certain configuration.

Near the critical points, designating phase transitions between states $(m',n')$ and $(m'',n'')$ with $ E^{0}_{(m',n')}= E^{0}_{(m'',n'')}$, the low temperature limit yields
\begin{equation}
    Y_{pair} \approx g_{(m',n')} e^{-\beta E^{0}_{(m',n')}} +g_{(m'',n'')} e^{-\beta E^{0}_{(m'',n'')}} =(g_{(m',n')} +g_{(m'',n'')})  e^{-\beta E^{0}_{(m',n')}}
\end{equation}
and
\begin{equation}
    \frac{\partial Y_{pair}}{\partial \beta} \approx -(g_{(m',n')} +g_{(m'',n'')}) E^{0}_{(m',n')}  e^{-\beta E^{0}_{(m',n')}} \;,
\end{equation}
which imply
\begin{equation}
    s \approx \frac{1}{2}k_B \ln ( g_{(m',n')} + g_{(m'',n'')} )\;.
\end{equation}
Hence, the peaks observed in the $s$ versus $\mu$ plots are related to the degeneracies of the coexisting states.

\begin{figure}
    \centering
    \includegraphics[width=0.8\textwidth]{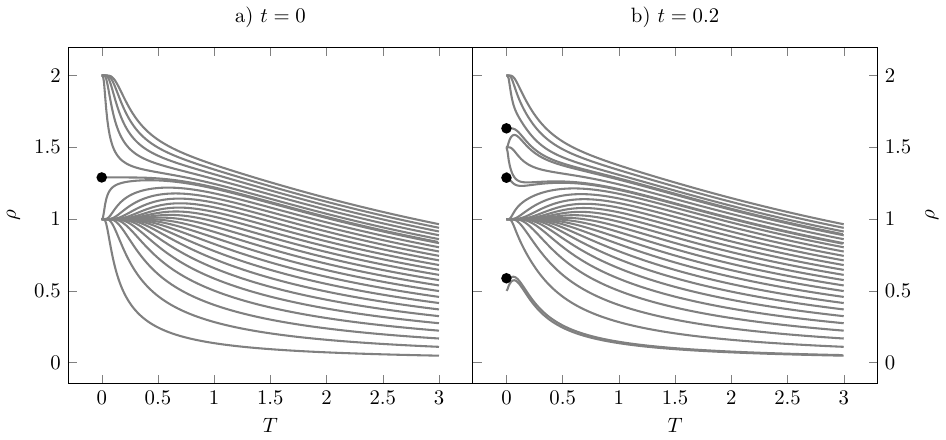}
    \caption{Density as a function of temperature for fixed pressures, ranging from $P = 0.15$ to $P = 5$ (bottom to top) at $U = 0$ and $V = 1$. Panel a) shows the atomic limit case $t = 0$, while in b) we consider $t = 0.2$. Black dots are the critical points shown in Figure~\ref{fig:2sc_U0_V1}: in a) we have $P_{c} \approx 4$, while in b) the critical pressures are $P_{c} \approx 0.16$, $3.92$ and $4.40$. The hopping slightly changes the critical pressure for the atomic limit anomaly, while introducing two additional ones. As $T \rightarrow 0$, all isobars collapse to specific densities, corresponding to the plateaus shown in panels c) and d) of Figure~\ref{fig:2sc_U1_V0.25}. The critical pressure corresponds to the chemical potential which generates the density transition at the ground state.}
    \label{fig:2sc_rho_x_T_const_P_U0_V1}
\end{figure}

The peak behavior shown in the residual entropy is known to be connected to a density anomaly in classical and bosonic systems~\cite{da_silva_residual_2015,rizzatti_waterlike_2019}. We show in Figure~\ref{fig:2sc_rho_x_T_const_P_U0_V1} the density at a fixed pressure for this system, which demonstrates a fermionic system can also display a density anomaly from a competition between $\mu$, $U$ and $V$. Indeed, for pressures slightly below the critical point (black dot), the density increases with temperature starting from $T=0$ until it reaches a maximum, at the TMD, and decreases as a regular fluid. Note that the critical pressures, related to the coexistence of states at $T=0$, also generate the entropy peaks of the  insets depicted in panels e) and f) of Figure~\ref{fig:2sc_U0_V1}. The pressure is kept constant using the relation $\Omega = -P \mathcal V$, where $\Omega$ is the Grand Potential, $P$ is the pressure and $\mathcal V$ is the volume of the system.

\begin{figure}
    \centering
    \includegraphics[width=0.8\textwidth]{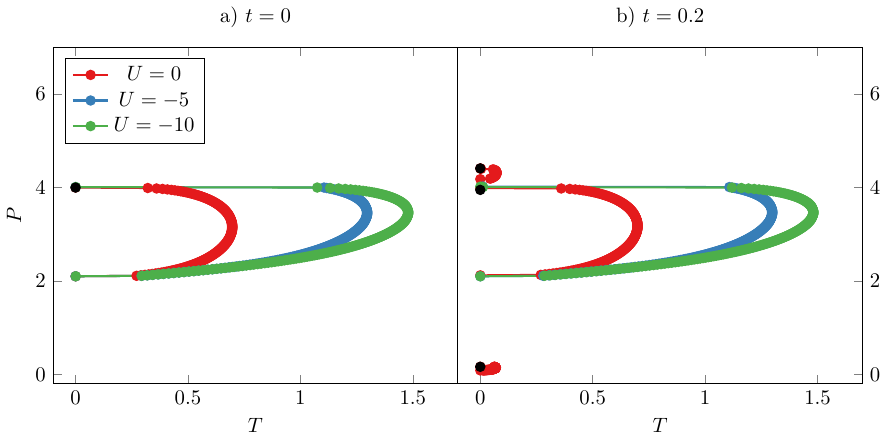}
    \caption{Temperature of maximum density (TMD) lines on the $PT$ plane at $V=1$ and different values of $U$. Panel a) shows the atomic limit case $t = 0$, while in b) we consider $t = 0.2$. The same color scheme is used in both plots. The black dots represent the critical point for the $U = 0$ case, in connection with the black dots of Figure~\ref{fig:2sc_rho_x_T_const_P_U0_V1}. A single density anomaly region was found for the negative $U$ cases since the $(1, 0)$ and $(2, 1)$ phases vanish.}
    \label{fig:PT_diagram_attractiveU}
\end{figure}

We map the  evolution of the density anomaly by plotting the TMD lines on the $PT$ diagram as the on-site interaction is varied according to Figure~\ref{fig:PT_diagram_attractiveU}. As we increase the attraction between fermions of different spins, the anomaly goes to higher temperatures, both in the atomic limit and with the hopping term. Since the residual entropy exhibits two peaks in panels e) and f) of Figure~\ref{fig:2sc_U0_V1}, we only have one region of density anomaly for $t = 0$, while the hopping term adds two other anomaly regions at low temperatures. We tested with values up to $U = -50$ and still found the density anomaly, suggesting that it is present for any $U \leq 0$ if $V > 0$. For $V < 0$ there is no density anomaly since the attractive effects of $U$ and $V$ cancel out, generating no competition with $\mu$.

\subsection*{Case 2: $U > 0$}

\begin{figure}
    \centering
    \includegraphics[width=0.8\textwidth]{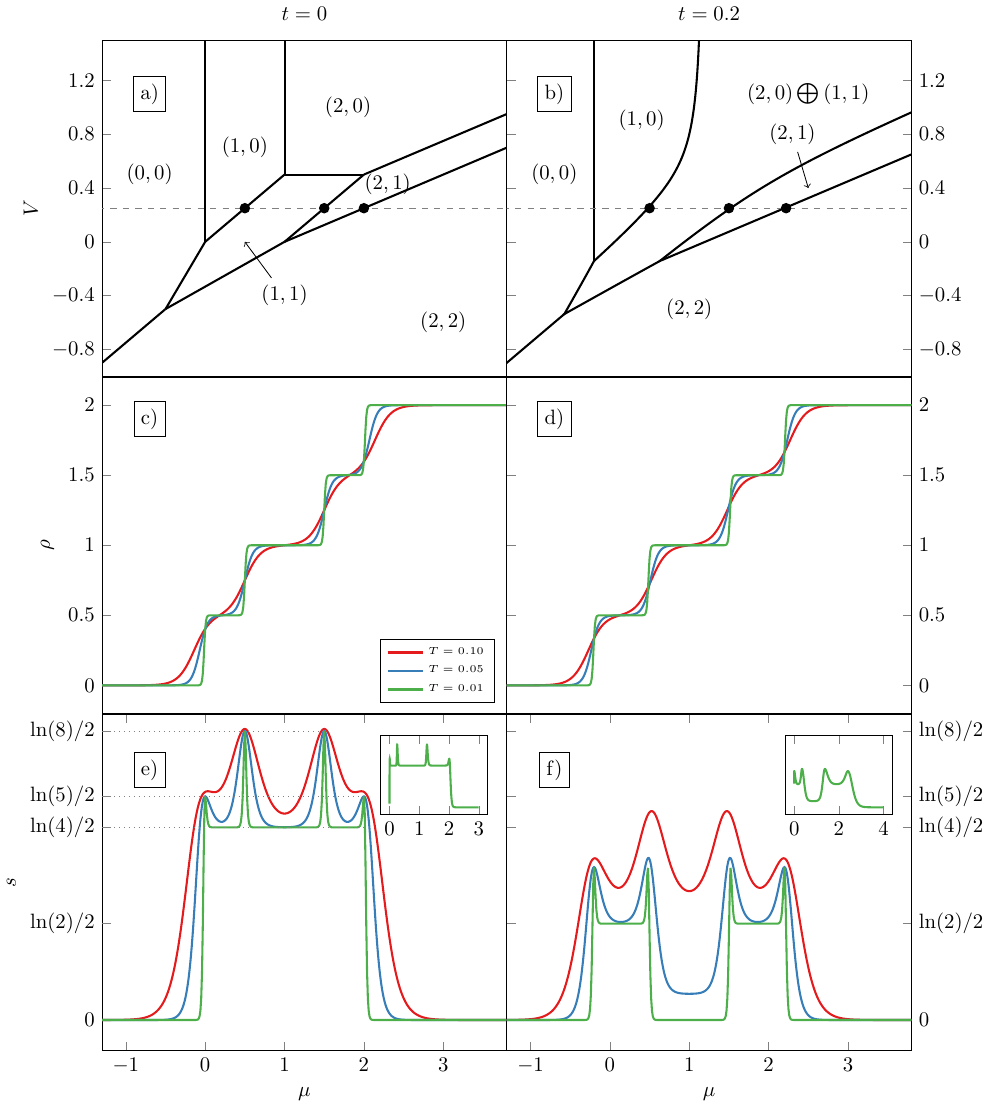}
    \caption{Panels a) and b) portray the ground state phase diagram at $U=1$. The black dots are explained in Figure~\ref{fig:2sc_rho_x_T_const_P_U1_V0.25}. Panels c) to f) show the density and entropy at a fixed value of $V = 0.25$ (dashed line) for various temperatures. The left column corresponds to the atomic limit $t=0$ while the right column refers to $t=0.2$. Insets in the last two panels are the entropy at $T = 0.01$ as a function of pressure. The same color scheme is used in all plots.}
    \label{fig:2sc_U1_V0.25}
\end{figure}

The ground state phase diagram for $U>0$ is shown in panels a) and b) of Figure~\ref{fig:2sc_U1_V0.25}. In this scenario, the system can organize itself into more phases in the atomic limit, yielding a richer phase diagram. As $U$ becomes larger, the size of the $(1, 0)$, $(2, 1)$ and $(1, 1)$ phases increase. All phase transitions in this diagram are first-order (discontinuous).

Let us give an interpretation for these results. For $V>\frac{1}{2} U$, the observed states are $(0,0)$, $(1,0)$, $(2,0)$, $(2,1)$ and $(2,2)$ depending on the imposed chemical potential. Their respective densities are $\rho=0$, $0.5$, $1$, $1.5$ and $2$. The three anomalous regions arise due to such successive transitions in the density parameter: from $(1,0)$ to $(1,1)$, from $(1,1)$ to $(2,1)$, and from $(2,1)$ to $(2,2)$. Also, note that all states are observed in this regime except for $(1,1)$, because the on-site repulsion represents less energy cost than the inter-site repulsion $V$ (i.e., the particles prefer to be on the same site, when possible). If we decrease $V$ to $0<V<\frac{1}{2}U$, the observed phases are $(0,0)$, $(1,0)$, $(1,1)$, $(2,1)$ and $(2,2)$. Basically, now the state $(1,1)$ becomes more stable compared to $(2,0)$, since the inter-site repulsion became smaller. Therefore, three regions of density anomaly are still present. Next, considering small negative values of $-\frac{1}{2}U<V<0$, the phases $(1,0)$ and $(2,1)$ vanish, since the system prefers to form two-site pairs, rather than being singly occupied at each site. This is why the stable configurations are $(0,0)$, $(1,1)$, $(2,2)$. Hence only one anomalous region will remain, originated from the transition $(1,1)$ to $(2,2)$ . Finally, decreasing $V$ to $V<-\frac{1}{2}U$, the system will tend maximize the number of two-site pairs, where only the states are present $(0,0)$ and $(2,2)$. Then no anomaly is seen in this regime.

\begin{figure}
    \centering
    \includegraphics[width=0.8\textwidth]{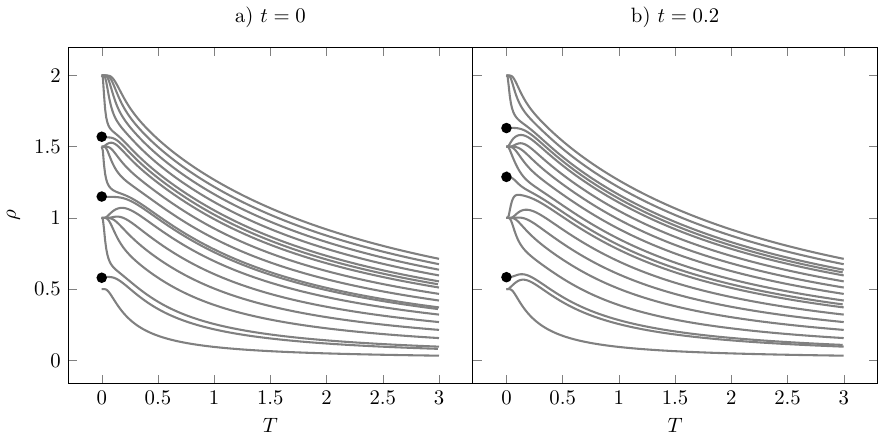}
    \caption{Density as a function of temperature with $U = 1$ and $V = 0.25$ for various pressures, ranging from $P = 0.01$ to $P = 3$ from bottom to top. Panel a) shows the atomic limit case $t = 0$, while in b) we consider $t = 0.2$. Black dots are the critical points shown in Figure~\ref{fig:2sc_U1_V0.25}: in a) we have $P_{c_1} \approx 0.25$, $P_{c_2} \approx 1.25$ and $P_{c_3} \approx 2$, while in b) the critical pressures are $P_{c_1} \approx 0.34$, $P_{c_2} \approx 1.38$ and $P_{c_3} \approx 2.39$.}
    \label{fig:2sc_rho_x_T_const_P_U1_V0.25}
\end{figure}

This multiplicity of states generates a considerable effect in the density as a function of temperature, as can be seen in Figure~\ref{fig:2sc_rho_x_T_const_P_U1_V0.25}: we now have multiple regions of density anomaly in the atomic limit, which are connected to the multiple peaks of the residual entropy, as shown by the insets of panels e) and f).

Regarding the ground state phase diagram with the inclusion of the hopping: the highly degenerate states $(1,1)$ and $(2,0)$ gives place to this non-degenerate configuration $(2,0)\bigoplus(1,1)$ which is a mixture of them, a linear combination of $\ket{\uparrow,\downarrow}, \ket{\downarrow,\uparrow}, \ket{\uparrow \downarrow,0}, \ket{0,\uparrow \downarrow}$. This doesn't mess with the anomalies because all these states have the same density $\rho=1$ (transitions between different densities aren't destroyed).

\begin{figure}
    \centering
    \includegraphics[width=0.8\textwidth]{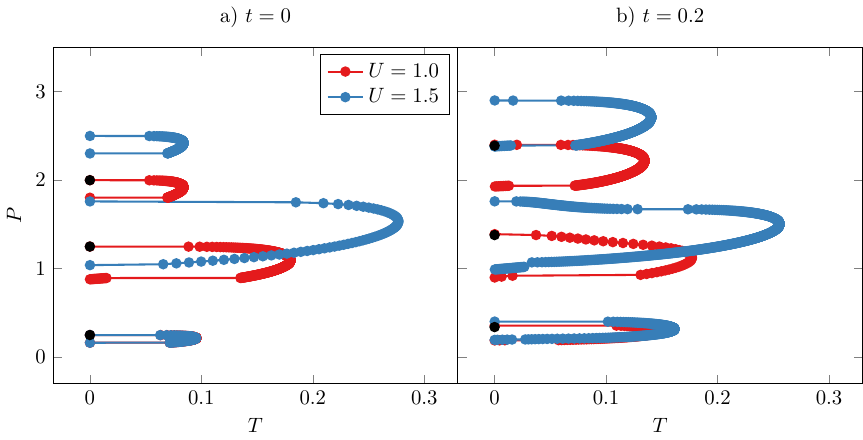}
    \caption{Temperature of maximum density (TMD) lines on the $PT$ plane at $V = 0.25$ for different values of $U$. Panel a) shows the atomic limit case $t = 0$, while b) considers a finite hopping of $t = 0.2$. Both panels use the same color scheme. The black dots represent the ground state phase transition for the $U = 1$ case, in connection with the black dots of Figure~\ref{fig:2sc_rho_x_T_const_P_U1_V0.25}.}
    \label{fig:PT_diagram_repulsiveU}
\end{figure}

In Figure~\ref{fig:PT_diagram_repulsiveU} we show the location of the density anomaly in a $(P, T)$ phase diagram, mapping the TMD curves. Making the on-site interaction more repulsive moves the upper anomalies to higher pressures and temperatures. The lowest-pressure anomaly is not affected by this change. As we increase $t$, the lower anomaly region decreases in size, going to regions of lower temperature and pressure. The higher region also vanishes. The intermediate transition grows to larger temperatures and pressures. We performed calculations up to $t = 50$ and verified the intermediate region is always present, indicating that this density anomaly region exists for any high value of $t$.

\begin{figure}
    \centering
    \includegraphics[width=0.8\textwidth]{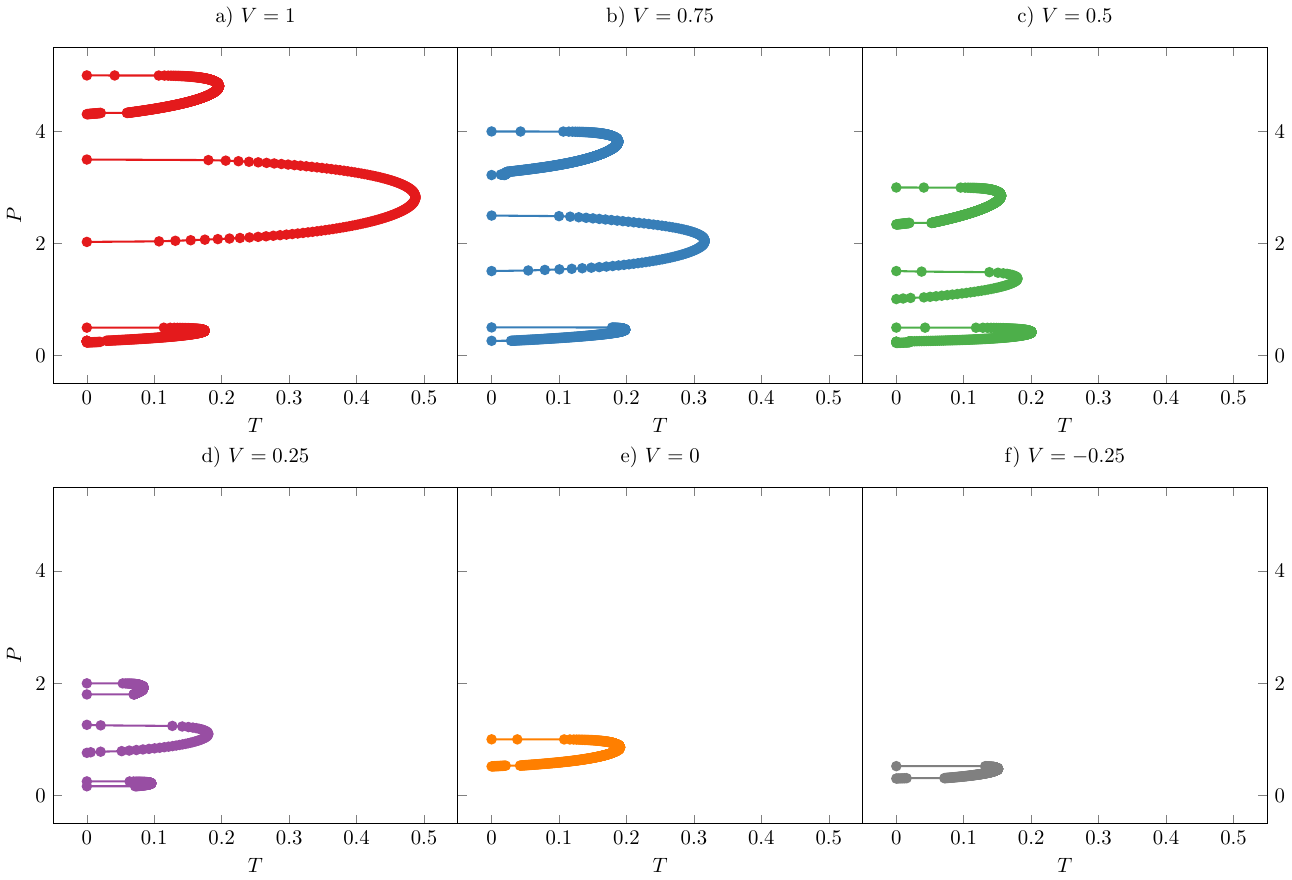}
    \caption{Temperature of maximum density (TMD) lines on the $PT$ plane for $U=1$ and different values of $V$ in the atomic limit. Note that the $P$ and $T$ axis have the same range for all plots. Dots are the calculated points and the solid lines are guides to the eye.}
    \label{fig:PT_diagram_repulsiveU_changeV}
\end{figure}

In the first case the behavior of the density anomaly as a function of $V$ was quite simple: it existed for $V > 0$. The current case requires a more detailed analysis to understand the multiple regions of density anomaly behave as $V$ is changed. This is shown in Figure~\ref{fig:PT_diagram_repulsiveU_changeV} for the atomic limit.
For a large positive $V$, all three density anomaly regions exists. As $V$ is lowered, the upper and lower regions shrink. The intermediate region is also reduced, vanishing smoothly.

\begin{figure}
    \centering
    \includegraphics[width=0.5\textwidth]{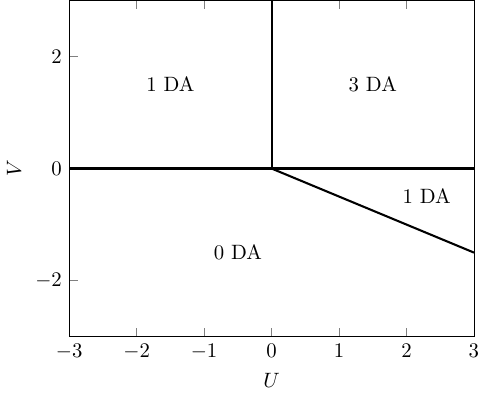}
    \caption{Number of density anomaly regions (DA) for given values of the parameters $U$ and $V$ in the atomic limit case.}
    \label{fig:UV-diagram}
\end{figure}

Lastly, we summarize our results with a $(U, V)$ diagram highlighting the existence of the density anomaly for given values of $U$ and $V$, for any pressure from $P = 0$ up to $P \rightarrow \infty$. This is shown in Figure~\ref{fig:UV-diagram}.

\section{Conclusions}
\label{sec:conclusion}
In this paper we studied the 1D extended Hubbard model in a pairwise approximation, which allows for an analytic solution of this complicated model considering a wide range of parameters. Through an analysis of its ground state and thermodynamic properties, we established a connection between phase transitions at zero temperature, residual entropies and the appearance of density anomalies. Depending on the on-site and first-neighbor interactions, we observed 0, 1 or even 3 regions of density anomaly on the phase diagram. Although in this kind of system there is no first-order line ending in a finite temperature critical point like in water, we show that this ground state phase transition propagates to higher temperature, generating a density anomaly for fermions.

Naturally, this pair approximation does not capture long range effects known to exist in the fixed density ground state phase diagram for this system, such as phase separation and superconductivity. The BEC superfluidity known to appear for very attractive values of $U$ is also not captured by this one dimensional approximation. Work to understand waterlike anomalies in these cases is underway.

We also note that, even in the atomic limit, our model does not approach the previously researched anomalies in the Bose-Hubbard model. Although both studies present similar behaviors as the repulsion is increased, a direct limiting procedure is not possible. This could be done using a Fermi-Bose-Hubbard model.

\section*{Acknowledgements}
We acknowledge that ChatGPT\footnote{\url{https://openai.com/product/chatgpt}} was used to review and enhance the Introduction chapter.
M.A.H. thanks CAPES for the PhD funding and M.A.O. Derós for the careful reading of this manuscript. E.O.R. thanks CNPq for the postdoc scholarship, grant 401867/2022-6.

\bibliographystyle{unsrt}
\bibliography{bibliography}

\end{document}